\documentclass[a4paper,10pt]{article}
\usepackage{a4wide}
\usepackage{cite}
\usepackage{amsmath,amsthm,amssymb,euscript,epsf,epsfig}
\usepackage{array}
\makeatletter
\renewcommand\section{\@startsection {section}{1}{\z@}%
                                   {-3.5ex \@plus -1ex \@minus -.2ex}
                                   {2.3ex \@plus.2ex}%
                                   {\normalfont\large\bfseries}}
\renewcommand\subsection{\@startsection{subsection}{2}{\z@}%
                                     {-3.25ex\@plus -1ex \@minus
                                     -.2ex}%
                                     {1.5ex \@plus .2ex}%
                                     {\normalfont\bfseries}}

\newcommand{\be}{\begin{equation}}
\newcommand{\ee}{\end{equation}}
\newcommand{\bea}{\begin{eqnarray}\displaystyle}
\newcommand{\eea}{\end{eqnarray}}

\makeatletter
\@addtoreset{equation}{section}
\makeatother

\begin{document}

\rightline{QMUL-PH-07-03} \rightline{CERN-PH-TH/2007-032}
\rightline{TIFR/TH/07-04} \rightline{PUPT-2223} \vspace{2truecm}

\centerline{\LARGE \bf Matrix Big Brunch} \vspace{1.3truecm}
\thispagestyle{empty} \centerline{
    {\large \bf J. Bedford${}^{a,b,}$}\footnote{E-mail address:
                                  {\tt j.a.p.bedford@qmul.ac.uk}},
    {\large \bf C. Papageorgakis${}^{c,}$}\footnote{E-mail address:
                                  {\tt costis@theory.tifr.res.in}},
    {\large \bf D. Rodr\'{\i}guez-G\'omez${}^{d,}$}\footnote{E-mail address:
                                  {\tt drodrigu@princeton.edu}}
    {\bf and}
    {\large \bf J. Ward${}^{a,b,}$}\footnote{E-mail address:
                                  {\tt j.ward@qmul.ac.uk}}
                                                       }

\vspace{.4cm}
\centerline{{\it ${}^a$ Centre for Research in String Theory, Department of Physics}}
\centerline{{ \it Queen Mary, University of London}} \centerline{{\it Mile End Road, London E1 4NS, UK}}

\vspace{.4cm}
\centerline{{\it ${}^b$ Department of Physics, CERN - Theory
    Division}}
\centerline{{\it 1211 Geneva 23, Switzerland}}

\vspace{.4cm}
\centerline{{\it ${}^c$ Department of Theoretical Physics, Tata
    Institute of Fundamental Research}} \centerline{{\it Homi Bhabha
    Road, Mumbai 400 005, India}}

\vspace{.4cm}
\centerline{{\it ${}^d$ Joseph Henry Laboratories, Princeton
    University, NJ 08544, USA}}

\vspace{2truecm}

\centerline{\bf ABSTRACT}
\vspace{.5truecm}

\noindent Following the holographic description of linear dilaton null
Cosmologies with a Big Bang in terms of Matrix String Theory put
forward by Craps, Sethi and Verlinde, we propose an extended
background describing a Universe including both Big Bang
and Big Crunch singularities. This belongs to a class of exact string backgrounds
 and is perturbative in the
string coupling  far away from the singularities, both
of which can be resolved using Matrix String Theory. We provide a simple
 theory capable of describing the complete evolution of this closed
Universe.

\newpage

\section{Introduction}

It is widely believed that string/M-theory should be singularity
free in its moduli space. So far the theory has succeeded in
resolving a number of static singular regions, but null and
space-like
 singularities still pose
a fundamental challenge. The standard Cosmological paradigm
assumes the existence of a Big Bang, out of which the Universe
originated. Indeed, its relic signature is the cosmic microwave
background radiation at a temperature of close to 2.7K which we
can measure today. Hence both from the purely theoretical and the
experimental points of view, the question of what happened at the
Big Bang is of extreme interest and any progress in the
understanding of non-static singularities is highly desirable.

An interesting step in this direction arising from string/M-theory
is a model proposed in \cite{CSV}.
By choosing a particularly simple background, namely a light-like
linear dilaton (LLD), Craps \textit{et al.} were able to obtain a
Matrix String Theory description \cite{BFSS,Motl,Banks,DVV} of the
early-time physics, which seems to be valid even at the Big
Bang-type singularity. One of the main attributes of the LLD
background is that the string worldsheet theory is still free,
since the dilaton does not contribute to the central charge of the
CFT. Therefore one would na\"ively expect that it should be
possible to make use of Matrix String Theory, by simply exchanging
$g_s$ for $g_s(x^+)$. Indeed, the authors of \cite{CSV} showed how
this
 guess turns out to be correct and obtained a resolution
of the singularity by explicitly deriving the matrix model via the
Discrete Light-Cone Quantisation (DLCQ) of  string theory in the
presence of an LLD and a sequence of
 dualities. Subsequently the one-loop matrix potential was calculated
 in \cite{craps2} and was found to be attractive while  vanishing at
 late times\footnote{An everywhere vanishing  answer was obtained in
   \cite{miaoli} although it is believed that this is due to
   calculating a time averaged version of the potential.}. These
 developments have generated great activity on
 closely related topics, \emph{e.g.}
 \cite{Das1,Das2,sethi,sethi2,She,miaoli2,miaoli3,bin,chenchen,ohta1}.
 Alternative holographic approaches to the  study of non-static Cosmological singularities
 involving the AdS/CFT correspondence have also been recently considered in the
 literature \cite{ho,linwen,trivedi1,trivedi2,bak,ohta3}. For a  comprehensive list of references
 on both types of constructions as well as on the  more general problem of
 resolving null and space-like singularities in string/M-theory we refer the reader to
 \cite{trivedi2} and the reviews \cite{costa1,craps}.

In this note, starting with the LLD background, we will construct
an extension to \cite{CSV} containing a Big Bang and a Big Crunch whilst
still having a perturbative matrix model description near the
singularities. This is motivated by the analysis of the LLD
Cosmology: It is well known that the linear dilaton background can
be obtained as a Penrose limit along a purely radially
incoming/outgoing geodesic\footnote{As we will see, a particular
sign for the dilaton corresponds to a choice between incoming or
outgoing geodesics.} of an NS5 background (see for example
\cite{HRV}). This suggests a possible modification of the LLD
background by taking the Penrose limit along an outgoing geodesic
and gluing it with the Penrose limit along an incoming geodesic at
the point $x^+=0$. The resulting background is simply flat space
with a light-like linear dilaton whose  sign is reversed in going
from $x^+<0$ to $x^+>0$.  Since in Einstein frame the dilaton acts
as a scale factor, this sign reversal implies the transition from
a Big Bang to a Big Crunch. The mathematical consistency of this
gluing procedure \cite{BI,Barrabes} requires the existence of a
non-trivial NS $B$-field supported only at the gluing junction.
As we will show, this background can also be obtained as a
solution to the IIA supergravity equations of motion.

For both the Big Bang and Big Crunch singularities, this solution
reduces to two copies of the LLD. The new ingredient of this construction is the
Cosmological turning point, $x^+=0$, at which the theory initially appears
to be singular. We study the supergravity description and
find a background in terms of
 a regulating parameter, which
once taken to zero leads back to the original solution. We will
argue that the generic features of \cite{CSV} are still valid and
that, in both the far past and far future, one can construct a mildly
modified version of Matrix String Theory describing both null
Cosmological singularities. Once again the interpretation of the turning
point is less transparent, but the physics appear to be captured by
perturbative strings. Hence it is still possible to write down a
Matrix String Theory valid even at the turning point.

The rest of this note is organised as follows: In Section 2 we
briefly review the Light-like Linear Dilaton background and the Matrix Big
Bang scenario of \cite{CSV}. In Section 3 we analyse the situation when viewed
as a Penrose limit of a non-extremal NS5 solution. We then go on
to describe the Big Bang-Big Crunch scenario as a gluing procedure in
Section 4. Section 5 is devoted to obtaining the same solution
from 10-d supergravity and providing a regularised
version before we discuss the Matrix String Theory description in
Section 6. We summarise and conclude in Section 7.

\section{Review of the Light-like Linear Dilaton background}

The LLD background is a remarkably simple time-dependent solution
of type IIA string theory involving flat Minkowski space in
co-ordinates $x^{\mu}=(x^+,x^-,\vec{x})$, with $\vec{x}$
representing the eight remaining space-like directions, and
string frame metric
\begin{eqnarray}
ds_{10}^2=-2dx^+dx^-+d\vec{x}^2\ .
\end{eqnarray}
There is also a dilaton given by $\phi=-Qx^+$, where $Q$ is a
constant. Flat space is still a string solution despite the
presence of  the light-like linear dilaton, since the latter, as
opposed to a space-like or time-like one, makes no contribution to
the conformal anomaly \cite{myers}.

By expressing the solution in Einstein frame, the metric is
rescaled by a factor of $e^{-\phi/2}$ giving
\begin{eqnarray}
\label{EF}
ds_E^2=e^{Qx^+/2}ds_{10}^2\ .
\end{eqnarray}
Cosmological evolution takes place in this frame. If one
interprets $x^+$ as the time variable, space-time originates
at a Big Bang as $x^+\rightarrow-\infty$ provided that $Q>0$.

It is not just that the metric vanishes as
$x^+\rightarrow-\infty$; it is clear that the string coupling
$g_s=e^{\phi}$ blows up in this limit and therefore we should
really be  thinking about M-theory rather than IIA strings.
Looking at the M-theory up-lift
\begin{eqnarray}
\label{uplift}
ds_{11}^2=e^{2Qx^+/3}ds_{10}^2+e^{-4Qx^+/3}(dx^{10})^2\ ,
\end{eqnarray}
with $x^{10}$ the eleventh direction, it is not hard to verify that there are divergent components of the Riemann tensor and that
the space-time is geodesically incomplete \cite{CSV}. The behaviour of the solution (\ref{EF}) is thus
that of a Cosmology with an initial Big Bang singularity in light-cone time.

In \cite{CSV}, Craps \textit{et al.} then go on to resolve this
singularity by going to a dual matrix model description in terms
of a 2-d super Yang Mills theory. This can be understood
either as the gauge theory defined on the cylinder with
a time dependent coupling, or as it having a constant coupling but
being defined on a time-dependent worldsheet given by the forward
quadrant of the  Milne orbifold of 2-d Minkowski space. In this
description the matrix model is weakly coupled when the string
theory is strongly coupled (\emph{i.e.} near the Big Bang
singularity), which renders it tractable. We will return to this
in Section 6.

The constant $Q$, in principle arbitrary, should be taken to be
positive if we are to interpret this solution as a Big Bang
Cosmology. By means of a boost $x^+\rightarrow \frac{x^+}{Q}$ and $x^-\rightarrow Qx^-$ we can always set the background to
\begin{eqnarray}
\label{SF}
ds_{10}^2=-2dx^+dx^-+d\vec{x}^2\ ,\quad \phi=-x^+\
\end{eqnarray}
\noindent in string frame or
\begin{eqnarray}
\label{EEFF}
ds_{E}^2=e^{x^+/2}\left(-2dx^+dx^-+d\vec{x}^2\right)\ ,\quad \phi=-x^+\
\end{eqnarray}
\noindent in Einstein frame, where we have a Big Bang singularity in the far past of
light-cone time $x^+$. The string coupling, $g_s=e^{-x^+}$ also diverges
at the singularity.

\section{Penrose limit of non-extremal NS5s and LLD}

It has been shown \cite{HRV} that the linear dilaton background can
be understood as a Penrose limit along a radial  null geodesic of
the solution for $N$ coincident  NS5-branes. Roughly speaking the
Penrose limit amounts to boosting to the speed of light while at
the same time  blowing up the neighbourhood of a given geodesic to
the whole space. Given that an observer falling freely into a
black hole cannot distinguish between the existence (or
non-existence) of horizons, it is to be expected that in both the
extremal and non-extremal NS5 cases the Penrose limit will yield
the same space. Indeed one can check explicitly that this is the
case, so for the sake of generality let us consider the near
horizon limit of the non-extremal NS5-brane solution \cite{HS}.
This is given by
\begin{equation}
ds^2=-\left(1-\frac{r_0^2}{r^2}\right)dt^2+d\vec{y}^2+\frac{Nl_s^2}{r^2}\Big\{\frac{dr^2}{\left(1-\frac{r_0^2}{r^2}\right)}+r^2\big(\cos^2\theta d\psi^2+d\theta^2+\sin^2\theta d\varphi^2\big)\Big\}\ ,
\end{equation}
\noindent where $\vec{y}$ corresponds to the five worldvolume co-ordinates of the NS5-branes, which are located at $r=0$.
The geometry in $\{t,r\}$ is that of a 2-d black hole whose horizon sits at $r=r_0$. There is also a non-zero dilaton given by
\begin{equation}
e^{2\phi}=\frac{\tilde{g}_s^2Nl^2_s}{r^2}\ ,
\end{equation}
\noindent where $\tilde{g}_s$ is the asymptotically weak string
coupling constant, and an NS 3-form field-strength given by
\begin{equation}
H_{(3)}=Nl^2_s\sin\theta\cos\theta\, d\psi\wedge d\theta\wedge
d\varphi\ .
\end{equation}
We will be interested in Penrose limits of this geometry along a
purely radial null geodesic. We closely follow the conventions of
\cite{HRV} and therefore parametrise the geodesic as
\begin{equation}
\left(\frac{dr}{du} \right)^2 = \frac{r^2}{Nl_s^2}\ ,
\hspace{1cm} dt=\frac{du}
{\Big(1-\frac{r_0^2}{r^2}\Big)}+dv\ .
\end{equation}
The first expression has the globally defined solution
\begin{equation}
\label{geodesics}
r=\bar{r} \exp \left(\frac{\pm u}{\sqrt{Nl_s^2}} \right)\ ,
\end{equation}
where the choice of sign corresponds to the choice between an
incoming or an outgoing geodesic. This follows from the fact that
if we compute the light-cone time velocity $\frac{dr}{du}$, this is either positive, and thus an outgoing geodesic, or
negative and thus an incoming geodesic. The parameter $\bar{r}$ is
an integration constant which sets the overall scale of the space.

We now introduce the following change of variables
\be
\begin{tabular}{c c c c}
$u=x^+\sqrt{Nl_s^2}$,&$v=\frac{x^-}{\sqrt{Nl_s^2}},\ $ &
$\theta=\frac{z}{\sqrt{Nl_s^2}},\ $ &
$\psi=\frac{w}{\sqrt{Nl_s^2}},\ $
\end{tabular}
\ee and after taking $N\rightarrow \infty$ one can easily see that
the metric reduces  to flat space in light-cone co-ordinates
whilst the 3-form field vanishes in this limit. Finally, the
dilaton becomes
\begin{equation}
\label{dd}
 e^{\phi}=\frac{\tilde{g}_s\sqrt{Nl_s^2}}{\bar{r}}e^{\mp
x^+}\ ,
\end{equation}
\noindent where we have implicitly assumed that the factor
$\frac{\tilde{g}_s\sqrt{Nl_s^2}}{\bar{r}}$ is positive and
we will also require it to be finite and small.
This is a reasonable assumption to make\footnote{We will come back
to this issue in Section 7.},  implemented by fine-tuning the
asymptotic string coupling $\tilde g_s$ to be weak and setting
the integration parameter $\bar r$ to scale with $N$.

From (\ref{geodesics}) we see that there are two possible solutions
for the radial geodesic corresponding to a particular choice of
sign. Let us choose the positive solution for definiteness which
gives $r=\bar{r}e^{x^+}$. After taking the Penrose limit (and
absorbing the constant factor in eq.(\ref{dd})) we obtain
 \be
 \label{LLD}
 ds^2=-2dx^+dx^-+d\vec{x}^2\ , \qquad \phi=-x^+\ ,
\ee \noindent which is precisely the linear dilaton background
that we found in (\ref{SF}) and (\ref{EEFF}) and was analysed in
\cite{CSV}\footnote{This alternative description and its relation to Little
String Theory  was also briefly touched on in that reference.}.
From this point of view, the Big Bang singularity at $x^+=
-\infty$ should be associated with $r = 0$, where the geodesic
along which we took the Penrose limit is hitting the NS5-branes. Had we
chosen the opposite sign, the geodesic would have hit the branes at
$x^+= \infty$, which from this point of view should correspond to
a Big Crunch singularity. Indeed, this picture can be verified by
repeating the geodesic completeness analysis of Section 2 for a
positive dilaton.

\section{The Big Brunch: Gluing the Big Bang to the Big Crunch}

Given that we have a way to construct a Big Bang and a Big Crunch, it
 is a natural to ask whether one can combine the two to obtain a
 closed Cosmology. The obvious way of doing that would be to pick a section of  each and glue them together to form a unique background involving both Big Bang and Big Crunch. Thus, let us choose  as a `geodesic'
 for the Penrose limit
\be r=\bar{r}e^{-x^+}\ \textrm{if}\ x^+> 0\ ,\qquad
r=\bar{r}e^{x^+}\ \textrm{if}\ x^+< 0\ , \ee \noindent  which
gives a dilaton \be \phi=x^+\ \textrm{if}\ x^+> 0\
,\qquad\phi=-x^+\ \textrm{if}\ x^+< 0\ . \ee

\noindent With this prescription we are gluing both geodesics at
$r=\bar{r}$. The background obtained in this way corresponds to
following geodesics going out of the NS5-branes at $x^+=
-\infty$, reaching a maximum point $\bar{r}$ at $x^+=0$ and then
coming back at $x^+=\infty$. Thus it should correspond to
evolution from a Big Bang to a Big Crunch. One could ask if the
reverse situation is also possible, namely, an evolution from a Big
Crunch to a Big Bang. As it will become clear later on, this
possibility is not allowed in terms of this particular
description because of the well-known positive energy theorems.

We have already seen that in the first patch the background is
described by  (\ref{LLD}), while in the second it is again (\ref{LLD}) but with
 the sign for the dilaton reversed. Because of their construction as
 a Penrose limit we know that both geometries are solutions to the
 supergravity equations of motion. However, nothing
 guarantees that the same will also hold for the gluing point. To find
 out whether
 our background indeed satisfies the supergravity equations of
 motion we first have to properly take care of the junction
 conditions. This set-up  could be thought of as a simple `cut-and-paste'
problem, which is well-known in gravity: In order to resolve it we
must ensure that both the metric and the extrinsic curvature are
continuous  functions when crossing the junction surface. If this
is not the case some extra matter  must be supplied at the
junction so that its stress-energy tensor compensates for the
discontinuity in the extrinsic curvature.

This procedure typically starts by choosing a vector normal to
 the gluing surface
and using it to compute the extrinsic curvature. In this case, however,
 we are gluing two copies of $\mathbb{R}^8$ at $x^+=0$. This
is a light-like surface, for which the normal vector declines into tangency
and the usual definition of the extrinsic curvature is no longer
valid. In turn, a slight variation of the algorithm was developed
 in  \cite{BI} to accommodate for this fact. We will follow the
latter prescription, \textit{i.e.} look for the `transverse' (in a
sense
 which will be described below)  extrinsic
curvature to the surface of gluing and, if the former is non-zero,
add some extra fields whose stress-energy tensor will  account
for the discontinuity. For this we need to revert to Einstein
frame.

 We wish to glue the following two patches
\begin{eqnarray}
\label{eyepatch}
ds_1^2&=&e^{x^+/2}(-2dx^+dx^-+d\vec{x}^2)\quad\,\,\,
\textrm{for}\,\, x^+ <0\nonumber \\
ds_2^2&=&e^{-x^+/2}(-2dx^+dx^-+d\vec{x}^2)\quad \textrm{for}\,\,
x^+
>0
\end{eqnarray}
\noindent at $x^+=0$. In order to proceed further we will make
use of the stress-energy tensor  $\tau$ as defined in
\cite{Barrabes} for a null shell, namely
\begin{equation}
8\pi\tau_{\mu\nu}=-[k_{\mu\nu}]+\frac{1}{2}\gamma l_{\mu}l_{\nu}\
,
\end{equation}
\noindent where
\begin{equation}
[k_{\mu\nu}]=-\frac{l^{\rho}}{2}\big(\partial_{\mu}g_{\nu\rho}+\partial_{\nu}g_{\mu\rho}\big)
\end{equation}
\noindent is the `extrinsic' curvature, as explicitly defined for the null case, and
\begin{equation}
\gamma
l_{\mu}l_{\nu}=[g^{\rho\lambda}\gamma_{\rho\lambda}]l_{\mu}l_{\nu}\
.
\end{equation}
The square braces stand for the following prescription
\begin{equation}
[F]=\lim_{x^+\rightarrow 0}\left(F_{x^+<0}-F_{x^+>0}\right)\ ,
\end{equation}
\noindent \textit{i.e.} $[F]$ is the difference between the initial
 and final  $F$ when crossing the junction. Note that this
expression knows about the ordering of the backgrounds in the
sense
 that we are evolving from $x^+<0$ to $x^+>0$. The
 vector  $l_{\mu}$ defines  the direction  tangent to the null surface
 along which we are gluing. For the complete details on the construction
 we refer the interested reader to \cite{Barrabes}.

By describing the gluing surface as the condition $f(x^+)=0$, we
obtain $l_{\mu}$ as
\begin{equation}
l_{\mu}=\partial_{\mu}f\ .
\end{equation}
\noindent Given that $f$ is just a function of $x^+$, it is clear
that the only non-zero entry for $l_\mu$ will be the one along
$x^+$. Indeed, it is easy to see that the only non-zero entry for
the stress-energy tensor at the junction is $\tau_{++}$. This turns out to be  constant and has
support on the gluing surface,  \emph{i.e.}
\begin{equation}
\label{T}
8\pi\tau_{++}=-4\delta(x^+)\ .
\end{equation}
In order to have a supergravity solution valid at the junction we
 need to  add some extra matter which will compensate for this
 difference. Motivated by the fact that our background comes from a
 certain limit of an NS5 geometry, where the only extra field is the
 NS $B$-field, we will  absorb (\ref{T}) through the introduction of such a field. Its stress-energy tensor, in
 a
 suitable normalisation for comparison with (\ref{T}), reads
\begin{equation}
T_{++}=-\frac{1}{32\pi}H_{+\mu\nu}H_+^{\phantom{+}\mu\nu}\ .
\end{equation}
\noindent At this point we will impose an extra condition,
assuming for the moment\footnote{We will provide good evidence in
support of this
  assumption in the following section.}
 that $H_{\mu\nu\rho}H^{\mu\nu\rho}=0$. This then leads to
\begin{equation}
\label{eq} H_{+\mu\nu}H_+^{\phantom{+}\mu\nu}=16\delta(x^+)\ .
\end{equation}
The background that we end up with in string frame then reads
\begin{eqnarray}
\label{stringystring} ds^2=-2dx^+dx^-+d\vec{x}^2\ ,\quad \phi =
|x^+|\ ,\quad H_{+\mu\nu}{H_{+}}^{\mu\nu}=16\delta(x^+)\ .
\end{eqnarray}
It is straightforward to  see that had we wanted to use the other ordering when
gluing the geodesics, namely the one which leads to a Big Crunch followed
by a Big Bang, we would have obtained the opposite sign in (\ref{T}). Therefore it would have been
impossible to account for that through the stress-energy tensor of any field,
for which the sign is fixed, or positive tension object. However this could
in principle be circumvented by the introduction of a new effective
field theory at the singularity, where string theory is strongly
coupled and it is expected that one will
have the appearance of new light degrees of freedom
\cite{seiberg1,seiberg2}, or by introducing negative tension objects
such as O-planes in the spirit of \cite{costa2}.

\section{Exploring the Supergravity picture}

 We have so far obtained a background by gluing  two
 geodesics along a null hyper-surface, in the process of engineering a
space-time incorporating both Big Bang and Big Crunch
singularities. The question now pertains to the supergravity
construction of such a solution. We begin with pure IIA
supergravity containing a non-zero $B$-field while keeping a flat
metric in string frame. We also have a
 light-like dilaton which is now a generic function of $x^+$. The
 equations of motion are given by
\begin{equation}
\label{einstein}
\frac{1}{4}\Big(\frac{1}{6}g_{\mu\nu}H^2-H_{\mu\alpha\beta}H_{\nu}^{\phantom
{\nu}\alpha\beta}\Big)+2\partial_{\mu}\partial_{\nu}\phi-g_{\mu\nu}\partial^2\phi=R_{\mu\nu}-\frac{1}{2}g_{\mu\nu}R\
,
\end{equation}
\begin{equation}
4(\partial\phi)^2-4\partial^2\phi+R+\frac{1}{12}H^2=0\ ,
\end{equation}
\begin{equation}
\partial_{\lambda}H^{\lambda}_{\phantom{\lambda}\mu\nu}-2\partial_{\lambda}
\phi H^{\lambda}_{\phantom{\lambda}\mu\nu}=0\ .
\end{equation}
Guided by the findings of the previous section, we will assume the
minimal form for the NS 3-form field strength, \emph{i.e.} the only
non-zero components of $H$ will be those with
$H_{+ij}=H_{+ij}(x^+)$, where $i$ and $j$ are indices labelling
the 8  spatial co-ordinates. All the equations of
motion are then trivially satisfied with the exception of
\begin{equation}
\label{eomB}
H_{+ij}{H_+}^{ij}=8\partial_+^2\phi\ .
\end{equation}
This simplicity arises because of the index structure of $H$ and the functional dependence of $H$ and $\phi$ only on $x^+$.
This means that $\partial^2\phi$, $(\partial\phi)^2$ and
$H^2=H_{\mu\nu\rho}H^{\mu\nu\rho}$ are all identically zero, therefore justifying the assumption  we made in the previous section \emph{a posteriori}. The
equation of motion for the $B$-field is satisfied for similar reasons. The only non-trivial terms come from the Einstein
equations which reduce to (\ref{eomB}) and show that the  dilaton acceleration is actually a source for the $B$-field.
Indeed, by assuming a simple linear dilaton solution  we get $H=0$, which brings us back to the case in \cite{CSV}.
Conversely had we set the $B$-field to zero, we would have obtained a
linear dilaton solution. Note that the acceleration of the dilaton is proportional to the square of $H_{+ij}$, and
therefore explicitly non-negative.

Ultimately the background solutions consistent with the equations
of motion are \be\label{ultimate} ds^2=-2dx^+dx^-+d\vec{x}^2\
,\qquad \phi=\phi(x^+)\ , \qquad H_{+ij}=H_{+ij}(x^+)\ , \ee where
the dilaton and the $B$-field are related as in (\ref{eomB}). The
3-form NS field-strength $H$ is given in terms of the 2-form
potential as usual, $H=dB$. However, as we require spatial
isotropy and homogeneity we will demand that
$H_{+ij}=\partial_+B_{ij}(x^+)$.

So far, we have constructed a set of supergravity solutions for
 a generic light-like  dilaton.
 Now, in very much the spirit of the background
constructed with the `cut-and-paste' procedure, we will
 take\footnote{Once again we have absorbed the asymptotic value of the dilaton in this definition.} the dilaton to be $\phi(x^+)=|x^+|$ since this
has the requisite linear behaviour in both domains. In this case
\begin{equation}
\label{modxeom}
H_{+ij}{H_+}^{ij}=16\delta(x^+)\ .
\end{equation}
The field $H_{+ij}^2$ is zero
everywhere apart from $x^+=0$. This is precisely the `cut-and-paste' solution as obtained from the
NS5-brane background (\ref{eq}), arising now in a more natural
way within the supergravity context. To summarise, the explicit solution
for our space-time is
\be
\begin{tabular}{c c c}
$ds^2=-2dx^+dx^-+d\vec{x}^2\ ,$ & $\phi=|x^+|\ ,$ &
$H_{+ij}^2=16\delta(x^+)\ .$
\end{tabular}
\end{equation}

We would like to point out that the above falls under a class of backgrounds  already considered in the literature \cite{HorowitzSteif,ohta2,panigrahi,ohta4}.  However in this note we concentrate on a particular example, namely the one representing the Big Bang-Big Crunch Cosmology.

\subsection{Supersymmetry Considerations}

Since the time-dependence of the problem has been introduced in
the form of light-cone time, we anticipate that our background
will preserve some fraction of supersymmetry. Here we will
explicitly show this by looking at the vanishing of the
supersymmetry variations relevant to the bosonic sector. These
yield \be
\begin{tabular}{c c}
$\delta
\psi_{\mu}=\Big(\partial_{\mu}-\frac{1}{8}\Gamma^{11}\Gamma^{\nu}\Gamma^{\rho}H_{\mu\nu\rho}\Big)\epsilon\
, $
&$\delta\lambda=\Big(\Gamma^+\partial_+\phi-\frac{1}{12}\Gamma^{11}\Gamma^+\Gamma^i\Gamma^jH_{+ij}\Big)\epsilon\
.$
\end{tabular}
\ee
The Clifford algebra relation  $\{\Gamma^+,\Gamma^i\}=0$ allows the
variation of the gaugino to  be re-written as
\begin{equation}
\Big(\partial_+\phi-\frac{1}{12}\Gamma^{11}\Gamma^i\Gamma^jH_{+ij}\Big)\times
\Gamma^+\epsilon=0\ .
\end{equation}
This holds, provided that $\epsilon$ is in the kernel of $\Gamma^+$.
If we now examine the variation of the gravitino, since $H$ has no $H_{-\mu\nu}$ component,
the only non-trivial equations are those for $\mu=+$ and $\mu=i$. For the
latter we find that the equation is of the form
\begin{equation}
\Big(\partial_i +  \frac{1}{8}\Gamma^{11}\Gamma^jH_{+ij}\Gamma^+\Big)\epsilon=0\ ,
\end{equation}
which is satisfied by decomposing  $\epsilon=f(x^+)\epsilon_0$
with $\epsilon_0$ a constant spinor.
We are thus left with the equation for $\mu=+$. This reads
\begin{equation}
\Big(\partial_+f-\frac{1}{8}\Gamma^{11}\Gamma^i\Gamma^jH_{+ij}f\Big)\epsilon_0=0\ .
\end{equation}
\noindent The solution is then given by
\begin{equation}
f=e^{\frac{1}{8}\Gamma^{11}\Gamma^i\Gamma^jB_{ij}}.
\end{equation}
Using the fact that
$\Gamma^+e^{\frac{1}{8}\Gamma^{11}\Gamma^i\Gamma^jB_{ij}}=e^{-\frac{1}{8}\Gamma^{11}\Gamma^i\Gamma^jB_{ij}}\Gamma^+$,
 the condition $\Gamma^+\epsilon=0$ gets translated
into $\Gamma^+\epsilon_0=0$ and we can thus conclude that the
solution preserves 16
 supersymmetries, \emph{i.e.} the same amount as in the linear
 dilaton case. We therefore have a family of $\frac{1}{2}$-BPS
 backgrounds with metric, dilaton and $B$-field as given by
 (\ref{ultimate}), with the additional constraint (\ref{eomB}).

\subsection{Regularising the Solution}

The supergravity background just constructed exhibits a sharp
 localisation at the turning point due to the delta function
 support for the $B$-field. In case this is of concern to the reader, we can smoothen out this singular behaviour
by constructing a regularised version in terms of a scalar parameter
$\epsilon$. We will then be able to take $\epsilon\rightarrow 0$ and
 recover the shell-like solution that we have thus far described.
Let us consider the case where we choose the $B$ field to lie
along two spatial directions such that $B_{12}$ is non-zero. This
enables us to forget about the tensorial character of $H$ for the
time-being (though it would be easy to consider a more generic
case) and allows us to define a regularised field strength
\begin{equation}
H_{+12}= \lim_{\epsilon \to 0} A\frac{\epsilon^a}{x^{+2}+\epsilon^2}\ .
\end{equation}
Here $A$ is a normalisation factor and $a$ a suitable exponent to be fixed
at a latter stage. The choice of this function is motivated by the fact that $H^2$ must be
a $\delta$-function, the regularised version of which is a  Lorentzian.

By using the equations of motion we can obtain the corresponding
dilaton. Since in the $\epsilon \rightarrow 0$ limit we should have an
absolute value, this allows us to fix both $a$ and $A$. A short calculation,
flat-space metric aside, reveals that the 3-form field-strength and  dilaton are given by
\be
\begin{tabular}{c c}
$\phi=\frac{2}{\pi}\Big(x^{+}\arctan
\left(\frac{x^{+}}{\epsilon}\right)+\epsilon \log
  \epsilon\Big)\ ,$ &
  $H_{+12}=\sqrt{\frac{32}{\pi}}\frac{\epsilon^{\frac{3}{2}}}{x^{+2}+\epsilon^2}\
  .$
\end{tabular}
\ee
\noindent It can be shown analytically that the dilaton goes to $\phi=|x^+|$
in the limit $\epsilon\rightarrow 0$. In the case of $H$ the
situation is more involved, however it can be proved that for a
well behaved function $f(x^+)$ (\emph{i.e.} finite as
$x^+\rightarrow \pm \infty$), as $\epsilon\rightarrow 0$ 
\begin{equation}
\frac{1}{16}\int^{\infty}_{-\infty}dx^+H_{+12}^2(x^+) f(x^+)=f(0)\
.
\end{equation}
Therefore, the regularised solution  exhibits the necessary
$\delta$-function behaviour and  reduces to the original one in that limit.

At this point we have constructed  a regularised version of the supergravity theory.
In fact we can go one step further and argue that this family of
backgrounds is also a
string theory solution to all orders in $\sigma$-model
perturbation theory. In \cite{HorowitzSteif} it was shown that a
certain class of backgrounds involving a metric, dilaton and NS
$B$-field do not receive higher
$\alpha'$ corrections. These include fields of
the form
\begin{eqnarray}
\label{allorder} ds^2 &=&
-dx^+dx^-+d\vec{x}^2+F(x^+,\vec{x})(dx^+)^2\nonumber\\
H_{\mu\nu\rho}&=&
A_{ij}(x^+)l_{[\mu}\nabla_{\nu}x^{i}\nabla_{\rho]}x^j\nonumber\\
\phi&=&\phi(x^+)\ ,
\end{eqnarray}
where $l^{\mu}$ is a null Killing vector encapsulating the fact
that the metric is independent of $x^-$. The above are then solutions to all
orders in $\alpha'$ if
\begin{eqnarray}
\label{mastereq}
\partial^2F+\frac{1}{18}A_{ij}A^{ij}+2\partial_{+}^2\phi=0\ .
\end{eqnarray}
Our solution has $F\equiv 0$, $\phi=|x^+|$ and $H_{+ij}=CA_{ij}$,
where $C$ is a constant, together with a suitable rescaling of
$x^+$ and $x^-$ to match the metrics. Up to some normalisation, the condition
(\ref{mastereq}) is thus the same as our equations of motion  (\ref{modxeom}).
 Hence, the Big Bang-Big Crunch
background that we have described belongs to the more general
family studied in \cite{HorowitzSteif} and is a solution
of string theory to all orders in $\alpha'$. It is important to
note that, because of the special $x^+$ dependence of the fields,
a similar statement also applies to the regularised version of the
solution. Since we  obtain the  shell-like background through a
path in parameter space which lies entirely inside this family of
exact string solutions, we feel confident that it is indeed a good
background for string propagation.

\section{A Theory describing the Big Bang-Big Crunch}

Having ensured that we have obtained a fully
consistent string background, we can start studying string
propagation. Consider the (light-cone) time-dependent effective string
coupling
\begin{equation}\label{gstime}
g_s(x^+)=g_0e^{|x^+|} \ ,
\end{equation}
where $g_0\equiv\frac{\tilde{g}_s\sqrt{Nl_s^2}}{\bar{r}}$. From this we
see that for $x^+\rightarrow \pm \infty$ the string coupling tends
to infinity and so a perturbative expansion in string loops does
not make sense.

Let us forget about the point $x^+=0$ for the moment. We are then
left with strings in an LLD background, which become strongly
coupled at the Big Bang or  the Big Crunch in each respective
patch. Thus, as proposed by Craps \emph{et al.}, we can conclude
that away from $x^+=0$ the full
 dynamics will be captured by a Matrix String Theory. The latter is
 described by an action
\be \label{S} S = \frac{1}{2\pi l_s^2} \int d^2\sigma \
\textrm{Tr} \left( \frac{1}{2}(D_\alpha X^i)^2 +
\theta^T\gamma^{\alpha}D_{\alpha}\theta + g_s^2 l_s^4 \pi^2
F_{\alpha\beta}^2
  - \frac{1}{4 \pi^2 g_s^2 l_s^4}[X^i,X^j]^2 + \frac{1}{2\pi g_s
l_s^2} \theta^T\gamma_i [X^i,\theta]\right)\ ,
 \ee
\noindent with the periodic identification $\sigma\sim \sigma + 2\pi
l_s$ and where the string coupling is given by (\ref{gstime}).
 The Yang-Mills coupling is identified with the inverse
product of the string length and the string coupling \be
g_{YM}\equiv \frac{1}{g_s l_s} \ee and it is obvious that this
gauge theory becomes weakly coupled when the string theory is
strongly coupled and vice-versa.

Proceeding in more detail along the lines of \cite{CSV}, we will
once again  assume a $B$-field whose only
non-vanishing
 element is $B_{12}$. In the usual Matrix String Theory construction in
flat space one considers  a light-like
compactification $x^-\sim x^-+2\pi R$, which is accompanied by a
small shift in $x^+$ due to the Sen-Seiberg argument \cite{sen,seiberg3}. However this  is no longer a symmetry in
this context due to the explicit dependence of the dilaton on that
co-ordinate. We can get around this problem by considering the
Lorentz transformation \be
\begin{tabular}{c c c}
$x^+=e X^+\ ,$ & $x^-=\frac{X^+}{2e}+\frac{X^-}{e}+\frac{X^1}{e}\
,$ & $x^9=X^++X^1\ ,$
\end{tabular}
\ee
\noindent in terms of which the original light-like  compactification
 $(x^+,x^-,x^9)\sim (x^+,x^-,x^9)+(0,R,e R)$ becomes just $X^1\sim
X^1+e R$, \textit{i.e.} is only a usual space-like
 compactification. Hence we can first T-dualise along the $X^1$ direction
 and then  perform an S-duality to end up with a system of
 coincident D1-branes wrapped on a spatial direction. The effective
 theory describing the latter is the sypersymmetric completion of the non-abelian DBI action 
  and it can be seen that it
 indeed reduces to its Yang-Mills truncation (\ref{S}) in
 the limit where $e\rightarrow 0 $ \cite{CSV,craps}.

 As it stands the above description is not valid at $x^+=0$, since at that
 point we also have the appearance of the NS $B$-field. However because
 the latter
 is of the form $B=B_{12}(x^+)dx^1\wedge dx^2$ it will remain unchanged
 under the Lorentz transformation. If we now perform the duality
 sequence, starting with the T-duality along $X^1$, we eventually
 arrive at
\begin{equation}
B \rightarrow\, C^{(2)}\ ,
\end{equation}
\noindent \emph{i.e.} we map the NS field into a RR 2-form
potential. This
 naturally couples to the worldvolume theory of the D1-strings.
Hence, the effect of $B$ will show up in the theory of
 IIA F1s as if it were a $C^{(2)}$ in the theory of IIB D1-strings,
 giving rise to a Chern-Simons term of the form
\begin{equation}
S_{CS}=\frac{1}{2\pi l_s^2}\int \ \textrm{Tr}\ \big( P[B]\big)\ .
\end{equation}
We conclude that if we add this extra term $S_{CS}$ to eq. (\ref{S}), we will obtain an action valid for all
 light-cone time, including $x^{+}=0$, the bosonic sector of which
 will be described by
\begin{equation}
\label{Sfull} S = \frac{1}{2\pi l_s^2} \int d^2\sigma \
\textrm{Tr} \Big( \frac{1}{2}(D_a X^i)^2 + g_s^2 l_s^4 \pi^2
F_{a b}^2- \frac{1}{4 \pi^2 g_s^2 l_s^4}[X^i,X^j]^2
+D_{a}X^1D_{b}X^2\epsilon^{a b}B_{12}\Big)\ .
\end{equation}

However in order to be confident about the validity of our
approach we should work with the regularised version of the
background and perform the duality sequence in the explicit
presence of the regulator. After going to the Lorentz transformed
co-ordinates adapted to the spatial compactification, one can check
that the background in string frame is:
\begin{equation}
ds^2=re^{-\Delta}\Big(-2dX^+dX^-+(dX^9)^2+d\vec{x}^2\Big)\ ,
\end{equation}
\noindent where
\begin{equation}
\Delta=\frac{2}{\pi}\left(eX^+\arctan\left(\frac{eX^+}{\epsilon}\right)+\epsilon\log\epsilon\right)\
.
\end{equation}
The dilaton becomes
\begin{equation}
\phi=-\Delta+\log \hat r\ ,
\end{equation}
where the parameter  $\hat r$  is related to the radius of the original
 light-like compactification as
\begin{equation}
\hat r=\frac{eR}{2\pi l_s}\ .
\end{equation}
Additionally the 3-form potential, which has become RR because of the S-duality, is given by
\begin{equation}
F_{+12}=\sqrt{\frac{32}{\pi}}\frac{\epsilon^{\frac{3}{2}}}{e^2(X^+)^2+\epsilon^2}\
.
\end{equation}

We can now consider the theory of  D1-strings. Following \cite{CSV}
 we will parametrise the action by choosing the following gauge
\be
\begin{tabular}{c c c}
$X^9=\frac{\sigma}{r}\ ,$ & $X^+=\frac{\tau}{r\sqrt{2}}\ ,$ & $X^-=\frac{\tau}{r\sqrt{2}}+\sqrt{2}y\ .$
\end{tabular}
\ee
\noindent
 Then, starting with the effective action for a single D1-brane we can
 plug in the ansatz and expand up to quadratic order in the fields. Once we
 take the gauge choice into account the action reduces to
\begin{equation}
S=\frac{1}{4\pi l_s^2}\int \, d^2\sigma\left[
(\partial_{\tau}\vec{x})^2+(\partial_{\tau}y)^2-(\partial_{\sigma}\vec{x})^2-(\partial_{\sigma}y)^2+\epsilon^{ab}\partial_{a}x^1\partial_bx^2B_{12}\right]\
.
\end{equation}
This is  the abelian version of (\ref{Sfull}) but with
$B$ replaced by its regularised counterpart. At the non-abelian
level, the
 action one would get would be nothing but (\ref{Sfull}) with the
 appropriately regularised fields. At this stage we can safely  take
the limit $\epsilon\rightarrow 0$ to exactly recover (\ref{Sfull}).

Given that the behaviour of our action mimics that of \cite{CSV},
it will be valid at both the Big Bang and Big Crunch
singularities. The main difference of our description lies in the
introduction of the $B$-field at $x^+=0$. Around that point the
string coupling constant is $\sim g_0$. By tuning this to be small
enough the commutator term in (\ref{Sfull}) is forced to vanish
leading  to the usual Green-Schwarz superstring in the presence of
an NS $B$-field.

\section{Conclusions and Outlook}

In this note we have extended the scenario presented in \cite{CSV}
to include a Cosmological evolution from a Big Bang to a Big
Crunch. We were able to do so by appropriately gluing together two
copies of the LLD background. The gluing procedure forced us to
introduce an extra NS $B$-field which was supported only at the
turning point, $x^+=0$. This particular solution was
also  seen as a certain representative of a class of supergravity
solutions involving the metric tensor, $B$-field and dilaton, which
is a family of exact string backgrounds. Since the former was also obtained as a limit of a
regularised solution belonging to the same class, we feel
confident
 about its validity despite the $\delta$-function support for $B$. After obtaining  the background involving
  the evolution from a Big Bang to a Big
 Crunch, we then proceeded along the lines of \cite{CSV} to construct
 a Matrix String Theory description of the physics.
This was shown to be valid at the Big Bang, the Big Crunch and also at
 the turning point.

Despite the above results, the physical interpretation of the
turning point is still quite unclear. This is closely related
to the somewhat exotic character of the  NS $B$-field
whose field-strength satisfies $dH=0$ and $d*H=0$. However we
argued that the turning point does indeed admit a perturbative
description: In terms of the NS5 picture, we ought to ensure that
\begin{equation}
g_0=\frac{\tilde{g}_s\sqrt{N l_s^2}}{\bar{r}}\ll 1\ ,
\end{equation}
 where $\tilde{g}_s$ is the usual asymptotic value for the
string coupling of the original NS5s, which is naturally taken to
be
 small so that  perturbative strings are well defined. Moreover it is easy
 to see that the
 curvature of the NS5 background
 is given by
\begin{equation}
\frac{1}{N}\left(1-\frac{4Nl_s^2}{6r^2}\right)\ .
\end{equation}
\noindent Since  we are interested in the limit where
$N\rightarrow \infty$, if we are to keep this small around $r\simeq\bar r$ we should require  that
\begin{equation}
\frac{Nl_s^2}{\bar r^2}\ll 1\ .
\end{equation}
This can be satisfied provided that $\bar r$ is large enough in
string units, and follows from our requirement that $\bar{r}$
scales with $N$. Given that the solution also remains under
control in the NS5-brane description, it could be possible to find
a correspondence between our set-up and 
the NS5 worldvolume theory \cite{sorokin} or Little String Theory
\cite{berkooz,seiberg4,aharony}. In this context it would be nice
to have a better understanding of the regularisation, which could
be related to a limiting procedure in the NS5 picture. However
even without referring to the NS5 interpretation, it seems clear
that it is possible to
 go to a corner of the  moduli
 space so that the theory is perturbative at the turning point. It
 would be interesting to perform a more detailed analysis of this
 point, in which the dual
 description in terms of the Milne orbifold
 worldsheet of \cite{CSV} could be of help. However all these issues
 are beyond the scope of
 this note, and we will leave them open for future investigations.

\vspace{1cm}

\textbf{Acknowledgements}

\vspace{0.5cm}

\noindent We would like to thank J. de Boer, B. Craps, B. Fiol, B.
Janssen, Y. Lozano, K. Narayan, S. Ramgoolam, \mbox{R. Roiban}, K.
Skenderis, S. Trivedi, M.A.Vazquez-Mozo and K. Zoubos for helpful
comments and discussions. C.P. is grateful to the Centre for
Research in String Theory at Queen Mary,
 University of London
for generous hospitality and support while this work was in progress and  to the people of India for supporting research in
String Theory. J.B. and J.W. would like to acknowledge Queen Mary
studentships and Marie Curie Early Stage Training grants. D.R-G. is supported by a Fulbright-MEC FU-2006-0740 fellowship. He would like to thank Amsterdam University for kind hospitality while this work was in progress.

\end{document}